\documentclass[a4paper,12pt]{article}
\usepackage[top=3cm,bottom=3.4cm,left=2cm,right=2cm]{geometry}

\usepackage[dvipdfmx]{graphicx}
\usepackage{bm}
\usepackage{amsmath}
\usepackage{amssymb}
\usepackage{float}
\usepackage{url}
\usepackage{slashed}

\newcommand{\nn}{\nonumber\\}

\renewcommand{\thepage}{}
\makeatletter
\@addtoreset{equation}{section}
\renewcommand{\theequation}{\thesection.\@arabic\c@equation}
\makeatother
\renewcommand{\thefootnote}{\fnsymbol{footnote}}
\begin{document}
\begin{titlepage}
\title{
\vspace*{-4ex}
\hfill
\begin{minipage}{3.5cm}
\end{minipage}\\
 \bf 
Supersymmetric nonlinear sigma models as anomalous gauge theories
\vspace{0.5em}
}

\author{Aya {\sc KONDO} and 
Tomohiko {\sc TAKAHASHI}
\\
\vspace{0.5ex}\\
{\it Department of Physics, Nara Women's University,}\\
{\it Nara 630-8506, Japan}}
\date{\today}
\maketitle

%
\vspace{7ex}

\begin{abstract}
\normalsize
We revisit supersymmetric nonlinear sigma models on the target manifold
$CP^{N-1}$ and $SO(N)/SO(N-2)\times U(1)$ in four dimensions.  These
models are formulated as gauged linear models, but it is indicated that
the Wess-Zumino term should be added to the linear model since the
hidden local symmetry is anomalous.  Applying a procedure used for
quantization of anomalous gauge theories to the nonlinear models, we
determine the form of the Wess-Zumino term, by which a global symmetry
in the linear model becomes smaller in the action than the conventional
one.  Moreover, we analyze the resulting linear model in the $1/N$
leading order. Consequently, we find that the model has a critical
coupling constant similar to bosonic models. In the weak coupling
regime, the $U(1)$ local symmetry is broken but supersymmetry is never
broken. In contrast to the bosonic case, it is impossible to find stable
vacua in the strong coupling regime as far as in the $1/N$ leading
order.  These results are straightforwardly generalized to the case of
the hermitian symmetric space.
\end{abstract}
\end{titlepage}

\renewcommand{\thepage}{\arabic{page}}
\renewcommand{\thefootnote}{\arabic{footnote}}
\setcounter{page}{1}
\setcounter{footnote}{0}
%
\tableofcontents

\section{Introduction}

A nonlinear sigma model is regarded as a low energy effective field
theory, where the relevant degrees of freedoms are massless
Nambu-Goldstone (NG) bosons associated with broken global symmetries.
Interestingly, any nonlinear sigma model based on the coset manifold is
gauge equivalent to a linear model with a so-called hidden local
symmetry (see \cite{Bando:1987br} and references cited
therein). Although the gauge fields for the hidden local symmetry are
redundant variables, dynamical vector bosons may be generated by quantum
corrections even in four dimensions.

In supersymmetric field theories, Zumino first recognized that the
scalar fields of nonlinear models take their values in a K\"ahler
manifold and gave an explicit form of the action for the Grassmann
manifold \cite{Zumino:1979et}.  More general nonlinear realization for
more general coset spaces was extensively studied in
\cite{Bando:1984cc,Itoh:1985ha,Itoh:1985jz,
Kugo:1983ma,Ovrut:1981wa,Lerche:1983qa,Lerche:1984ub,Lerche:1985hy}
and general methods to construct a nonlinear Lagrangian are provided.
The characteristic feature is that massless fermions appear as
supersymmetric partners of NG bosons. These NG bosons and their
fermionic partners are described by chiral superfields in four
dimensions with $N=1$ supersymmetry. Then, the target space must be the
K\"ahler manifold since chiral superfields are complex.

Supersymmetric nonlinear sigma models with hidden local symmetries were
studied on some K\"ahler manifolds in
\cite{Aoyama:1979zj,Lindstrom:1983rt,Hitchin:1986ea,rf:KugoSoken} and
then have been generalized by Higashijima-Nitta about twenty years ago
\cite{Higashijima:1999ki}. They showed that a supersymmetric nonlinear
sigma model is formulated as a linear gauge theory, if its target
manifold is the hermitian symmetric space.  However, importantly, this
is a classical correspondence between both models.

Supersymmetric nonlinear sigma models were studied in quantum field
theories and many interesting results have been revealed in two
dimensions
\cite{DiVecchia:1977nxl,Witten:1977xn,Witten:1978bc,DAdda:1978dle}.
However, nonlinear sigma models are nonrenormalizable in four
dimensions. So they are defined by the theory with ultra-violet momentum
cutoff as well as Nambu-Jona-Lasinio (NJL) model \cite{rf:NJL}, or by
some other non-perturbative methods.  Although supersymmetry increases
difficulties in analyzing the quantum dynamics, they seem not to be
physical but to be technical, similar to an ambiguity of subtraction in
NJL model, and so a relatively tractable problem.

Most crucially, a hidden local symmetry is generically anomalous in
supersymmetric nonlinear models in four dimensions, since the symmetry
acts on chiral superfields.  For example, let us consider the following
K\"ahler potential as a gauged linear model:
\begin{align}
 K(\phi,\phi^\dagger)=\phi^\dagger e^{2V} \phi- \frac{2}{g^2} V,
\nonumber
\end{align}
where $\phi_i\ (i=1,\cdots,N)$ is a chiral superfield and $V$ is a $U(1)$
gauge vector superfield.  The last term is a Fayet-Illiopoulos (FI) term
with a coupling constant $g$. The model has
the global symmetry $SU(N)$ and the local one $U(1)$.
In order to see this model to be equivalent to the $CP^{N-1}$ model,
it has been thought that one has only to take $\phi_N=1$ as a gauge
fixing condition \cite{rf:KugoSoken,Higashijima:1999ki}. Eliminating $V$
by the equation of motion, one may found the K\"ahler potential of
the $CP^{N-1}$ model, the target manifold of which is parameterized by
the remaining chiral superfields.
However, the important point is that the anomalous hidden local symmetry
does not allow us to take arbitrary gauge fixing condition. In this example,
$U(1)$ is anomalous and so it is impossible to transform to the
$CP^{N-1}$ model.

For one thing, we can avoid the anomaly problem by considering
non-anomalous hidden local
symmetries in the gauged linear model. Alternatively, one can add
additional chiral superfields coupled to the vector superfield in order
to cancel the anomaly. However, both methods are not helpful for
formulating the nonlinear sigma model based on the hermitian symmetric
space. 

In this paper, we will start with the supersymmetric nonlinear sigma
model, which includes only the chiral superfields and so is a
well-defined theory without the anomaly. Then, we will rewrite the model
by introducing an auxiliary vector superfield and performing a Legendre
transformation.  At this stage, the vector superfield is not a gauge
field since the original Lagrangian is not gauge invariant and the path
integral measure is not divided by the gauge volume. Next, we will
insert the Fadeev-Popov determinant to the partition function by
following the technique used for the quantization of anomalous gauge
theories in~\cite{Harada:1986wb}, which is an extension of the method of
\cite{Faddeev:1986pc}. As a result, we obtain the gauged linear model
with a Wess-Zumino term which is equivalent to the original nonlinear
sigma model.

We should comment that the conceptual setting of the above procedure is
not new, because it is almost the same strategy described by de~Wit and
Grisaru more than thirty years ago \cite{deWit:1985bn}.  In the case of
the $CP^{N-1}$ model, the chiral superfields $\phi^i$ include a
compensating field. They showed that the anomaly can be always
eliminated by adding local counter terms constructed by using the
compensator.  However, an advantage of our procedure is that it is
obvious which field is a compensator, while there are various options in
their arguments.  Consequently, a Wess-Zumino term can be uniquely
determined in our procedure.

We will explicitly deal with $CP^{N-1}$ and $SO(N)/SO(N-1)\times U(1)$
models, but our results can be generalized straightforwardly to other
target manifolds, because these models capture typical features of the
models without or with F-term constraint\cite{Higashijima:1999ki}. Both
nonlinear models will be formulated as anomalous gauged linear
models. Importantly, the symmetry of the action in the gauged linear
model is different from a conventional symmetry due to the effect of the
Wess-Zumino term. For instance, we will show that the action of the
gauged linear model for the $CP^{N-1}$ model has the symmetry
$SU(N-1)_{\rm global}\times U(1)_{\rm local}$, which is smaller than the
conventional symmetry $SU(N)_{\rm global}\times U(1)_{\rm local}$.  This
is essentially the same result as pointed out by de~Wit and Grisaru in
the discussion of anomalies and compensators \cite{deWit:1985bn}.

This paper is organized as follows. First, we will show the details
about the supersymmetric $CP^{N-1}$ model. In the section
\ref{sec:model}, we will explain the quantum equivalence between this
model and an anomalous gauged linear model with a Wess-Zumino term,
which is derived from the Jacobian factor for chiral superfields. In the
section \ref{sec:anomaly}, we will calculate a three-point vertex
function given by triangle diagrams and exactly determine the form of
the Wess-Zumino term in the theory including the momentum cut-off
$\Lambda$. For renormalizable theories, the Feynman integral for the
triangle diagram is expanded by the powers of $1/\Lambda$ and only
finite terms for $\Lambda\rightarrow \infty$ contribute to the anomaly
\cite{rf:Weinberg}.  Here, we will provide an exact anomalous term
depending on $\Lambda$, which includes higher power terms of
$1/\Lambda$.  In the section \ref{sec:global}, we will discuss that our
model is defined on the whole $CP^{N-1}$ manifold.
In the section \ref{sec:CPNpot}, we will analyze the
effective potential of the linear model in the $1/N$ leading order. We
find that the model has the critical coupling, below which the
$U(1)_{\rm local}$ symmetry is broken and supersymmetry is
unbroken. Remarkably, in contrast to the bosonic $CP^{N-1}$ model
\cite{Bando:1987br}, we will show that there is no stable vacuum beyond
the critical coupling in the $1/N$ leading order.  In the section
\ref{sec:vector}, we will discuss the vector supermultiplet which is
dynamically generated but unstable as similar to the bosonic
case\cite{Bando:1987br}. Interestingly, we observe that, when
approaching the critical point, the vector multiplet tends to become
massless. This behavior suggests the possibility that the $U(1)_{\rm
local}$ symmetry is restored at the critical coupling. Next, we will
consider $SO(N)/SO(N-2)\times U(1)$ model in the section
\ref{sec:SOmodel} and \ref{sec:SOpot} as an example of the nonlinear
model with F-term constraint. Although an F-term is added to the model,
the qualitative features are unchanged. Finally, we will give concluding
remarks in the section \ref{sec:conclu}. In the appendix
\ref{sec:appendix}, we present details of calculation of Feynman
integrals in the cut-off theory.

\section{Supersymmetric $CP^{N-1}$ model}

\subsection{Anomalous gauged linear models
\label{sec:model}}

The supersymmetric $CP^{N-1}$ model is defined by the Lagrangian
\begin{align}
 {\cal L}=\int d^2 \theta d^2\bar{\theta} K_0(\varphi,\,\varphi^\dagger),
\label{eq:LagCP}
\end{align}
where $\varphi_i\ (i=1,\cdots,N-1)$ are chiral superfields and $K_0$ is the
K\"ahler potential given by
\begin{align}
 K_0(\varphi,\,\varphi^\dagger)=\frac{1}{g^2} \log\Big(\frac{1}{g^2}
+\varphi^\dagger \varphi
\Big).
\label{eq:Kphi}
\end{align}
As well-known, this K\"ahler potential provides the Fubini-Study
metric for $CP^{N-1}$ manifold, which is parameterized by the complex fields
$\varphi_i$, $\varphi^*_i$. The parameter $g$ is a coupling
constant with the dimension of mass inverse.
The K\"ahler potential can be expanded at $\varphi=0$ as
\begin{align}
 K_0(\varphi,\,\varphi^\dagger)=\frac{1}{g^2}\log \frac{1}{g^2}
 +\varphi^\dagger \varphi
-\frac{g^2}{2}(\varphi^\dagger \varphi)^2+\cdots,
\label{eq:Kphiseries}
\end{align}
where the first term has no effect on the Lagrangian,
and so we find that the chiral field $\varphi$ is canonically normalized
in (\ref{eq:Kphi}).

By introducing an auxiliary vector superfield $V$, we can change the
K\"ahler potential into
\begin{align}
 K_0'(\varphi,\,\varphi^\dagger,\,V)=
e^{2V} \Big(\frac{1}{g^2}+\varphi^\dagger \varphi\Big)
-\frac{2}{g^2} V,
\label{eq:KphiV}
\end{align}
where the last term is a FI D-term.
The equation of motion of $V$ leads to
\begin{align}
 \frac{\delta K_0'}{\delta V}=2 e^{2V}
\Big(\frac{1}{g^2}+\varphi^\dagger \varphi\Big)-\frac{2}{g^2} =0
~~\Rightarrow -2V= \log\frac{1/g^2+\varphi^\dagger \varphi}{1/g^2}.
\end{align}
Substituting this back into (\ref{eq:KphiV}), we obtain the same
K\"ahler potential (\ref{eq:Kphi}) for the $CP^{N-1}$ model
up to irrelevant constant terms.

In (\ref{eq:KphiV}), we perform change of
variables such as
\begin{align}
 2V &\ \rightarrow\  2V-i(\lambda-\bar{\lambda})
\\
\varphi_i &\ \rightarrow\  e^{i\lambda}\varphi_i,
\label{eq:phichange}
\\
\bar{\varphi}_i &\ \rightarrow\  e^{-i\bar{\lambda}}\bar{\varphi}_i,
\label{eq:phibarchange}
\end{align}
where $\lambda$ is a chiral superfield. Then, we find the K\"ahler
potential to become
\begin{align}
 K(\phi,\,\phi^\dagger, V)
=\phi^\dagger e^{2V}\phi-\frac{2}{g^2} V,
\label{eq:Klinear}
\end{align}
where $\phi_i\ (i=1,\cdots,N)$ are chiral superfields:
\begin{align}
 \phi_i=\varphi_i\ (i=1,\cdots,\,N-1),~~~\phi_N=\frac{1}{g}\,e^{-i\lambda}.
\label{eq:KphiV2}
\end{align}
This K\"ahler potential gives a gauged linear model with the global
symmetry $SU(N)$ and the local symmetry $U(1)_{\rm local}$. If we take
$\phi_N=1/g$ as a gauge fixing condition for $U(1)_{\rm local}$,
the K\"ahler potential (\ref{eq:Klinear}) reproduces the expression
(\ref{eq:KphiV}) and then the first one (\ref{eq:Kphi}) by eliminating
$V$. Hence, it was 
claimed that the supersymmetric $CP^{N-1}$ model can be obtained from a gauged
linear model.

However, it should be noticed that $U(1)_{\rm local}$ is
an anomalous symmetry and this anomaly is
an obstruction in proving the equivalence between both models.
In order to include the anomaly, we have to deal with contributions from
path integral measures.
The idea is basically same as the
quantization of anomalous gauge theory\cite{Harada:1986wb}, 
although the original Lagrangian (\ref{eq:LagCP}) is not gauge
invariant in our case.

At first, we introduce the auxiliary vector superfield $V$
to the partition function of the $CP^{N-1}$ model:
\begin{align}
 Z=\int d\varphi d\varphi^\dagger \exp\Big(i\int d^8 z
  K_0(\varphi,\,\varphi^\dagger)\Big)
=\int d\varphi d\varphi^\dagger dV \exp\Big(i\int d^8 z
  K_0'(\varphi,\,\varphi^\dagger,\,V)\Big),
\label{eq:partfunc1}
\end{align}
where the superspace coordinate is denoted by
$z=(x,\theta,\bar{\theta})$, integration measures by $d^8 z=d^4 x
d^2\theta d^2\bar{\theta}$.
In general, the $V$ integration leads to not only $K_0$ as a saddle
point, but also higher order quantum corrections. However, in
supersymmetric theories, we have no quantum
corrections as proved by Higashijima-Nitta\cite{Higashijima:1999js} and
so this is an exact rewriting.

Let us define the Fadeev-Popov determinant $\Delta_f[V]$ for the gauge fixing
condition $f[V]=0$:
\begin{align}
 \Delta_f[V]\int d\lambda
  d\bar{\lambda}\,\delta\left(f[V^{(\lambda,\bar{\lambda})}]\right)=1,
\label{eq:FPdet}
\end{align}
where $d\lambda d\bar{\lambda}$ is a gauge invariant measure and
$V^{(\lambda,\bar{\lambda})}$ is a gauge transformation of $V$:
\begin{align}
 2V^{(\lambda,\bar{\lambda})}=2V+i(\lambda-\bar{\lambda}).
\end{align}
Inserting (\ref{eq:FPdet}) into (\ref{eq:partfunc1})
and changing an integration variable as
$V\,\rightarrow\,V^{(-\lambda,-\bar{\lambda})}$,
the partition function (\ref{eq:partfunc1}) is expressed in terms of the
functional integral over $\lambda$, $\bar{\lambda}$ and the
original fields:
\begin{align}
 Z&=
\int
  d\varphi\,d\varphi^\dagger\,{\cal D}V\,d\lambda\,d\bar{\lambda}\,
\exp\Big(i\int d^8z 
  K'(\varphi,\,\varphi^\dagger,\,\lambda,\,\bar{\lambda},\,V)\Big),
\\
{\cal D}V&\equiv dV \Delta_f[V]\,\delta(f[V]),
\end{align}
where $dV$ is assumed to be gauge invariant and so ${\cal D}V$
corresponds to a
gauge invariant measure divided by the gauge volume. The K\"ahler
potential $K'$ is given by
\begin{align}
&
  K'(\varphi,\,\varphi^\dagger,\,\lambda,\,\bar{\lambda},\,V)
=e^{2V}\Big\{\,\frac{1}{g^2}e^{i\bar{\lambda}}e^{-i\lambda}+
(\varphi^\dagger e^{i\bar{\lambda}})(e^{-i\lambda}\varphi)\Big\}
-\frac{2}{g^2}V.
\end{align}

If we take the chiral superfields $\varphi'=e^{-i\lambda}\varphi$
as integration variables, the functional
measure produces the Jacobian factor
derived from the relation \cite{KS,CPS}
\begin{align}
\frac{\delta \varphi'_j(z)}{\delta \varphi_k(z')} =\delta_j^k\,
\,e^{-i\lambda(z)}\frac{-\bar{D}^2}{4}\delta^8(z-z').
\label{eq:delphi}
\end{align}
Moreover, we change the variable from $\lambda$
to $\phi_N=e^{-i\lambda}/g$.
Since $\lambda$ is a chiral field, we have a similar
relation to (\ref{eq:delphi}):
\begin{align}
\frac{\delta \phi_N(z)}{\delta \lambda(z')}=
-i\,\frac{1}{g}\,e^{-i\lambda(z)}\frac{-\bar{D}^2}{4}\delta^8(z-z').
\label{eq:dellambda}
\end{align}
So, in the partition function integrated over the new variables, we
have the Wess-Zumino term with the factor $N$, in which $N-1$ and $1$
are coming from the measures of $\varphi_i$ and $\lambda$, respectively.
Finally, we can rewrite the partition function of the $CP^{N-1}$ model
as follows,
\begin{align}
&
 Z=\int d\phi\,d\phi^\dagger\,{\cal D}V\,
\,\exp\left(\,i \int d^8z\,K(\phi,\bar{\phi},V)+
i\,\alpha[V,\phi_N,\bar{\phi}_N]\,\right).
\label{eq:partfuncf2}
\\
&
\alpha[V,\phi_N,\bar{\phi}_N]
=-\frac{N}{16\pi^2}\int d^4 x d^2 \theta 
\log (g\,\phi_N)\,
 W^\alpha W_\alpha
+{\rm h.c.}+O(1/{\Lambda}^2),
\label{eq:alphaV}
\end{align}
where the K\"ahler potential is given by (\ref{eq:Klinear}).
$\alpha[V,\phi_N,\bar{\phi}_N]$ is the anomalous term generated by the
Jacobian factor.
$\Lambda$ is the ultraviolet cut-off parameter to regularize the
functional measure \cite{KS,CPS}, in which the leading term is given by
the Wess-Zumino term for $U(1)_{\rm local}$.

Consequently, we show that the supersymmetric $CP^{N-1}$ model is
quantumly equivalent to the theory given by the K\"ahler potential
(\ref{eq:Klinear}) and the F term (\ref{eq:alphaV}).  This $F$ term
reduces the flavor symmetry to $SU(N-1)$ and so the action of this
gauged linear model has the symmetry $SU(N-1)\times U(1)_{\rm local}$.

\subsection{Global structure and inhomogeneous coordinates
\label{sec:global}}

We have started from the action (\ref{eq:Kphi}) of the $CP^{N-1}$ model
and then have rewritten its partition function as that of the linear
model (\ref{eq:partfuncf2}). In the action (\ref{eq:Kphi}), $\varphi_i$
denote local affine coordinates of the $CP^{N-1}$ manifold and so this
coordinate patch does not cover $CP^{N-1}$.

First, let us reconfirm that the partition function given by the K\"ahler
potential (\ref{eq:Kphi}) is defined on the whole manifold, while the
action is represented by the local coordinates.  For simplicity, the
coupling constant is set to be one. $\varphi_i$ are local coordinates in
a patch, which is expressed by $U_0$. In the case of $\varphi_k\neq 0$,
we can introduce an affine coordinate system in the coordinate patch
$U_k$:
\begin{align}
 \varphi'_k=\frac{1}{\varphi_k},~~~\varphi_i'=\frac{\varphi_i}{\varphi_k}
~~~(i\neq k).
\label{eq:cdch}
\end{align}
Importantly, the $N$ coordinate patches $U_i\ (i=0,\cdots, N-1)$ cover
the $CP^{N-1}$ manifold.

Under the coordinate change (\ref{eq:cdch}), 
the K\"ahler potential
(\ref{eq:Kphi}) is transformed to
\begin{align}
 K_0(\varphi,\,\varphi^\dagger)= K_0(\varphi',\,{\varphi'}^\dagger)
+f(\varphi')+f^*({\varphi'}^*),
\end{align}
where $f(\varphi)$ is the holomorphic function $f(\varphi')=-\log
\varphi'_k$.  Since both holomorphic and antiholomorphic terms are
vanished in the action after supercoordinate integration, the action has
the same expression with respect to the coordinates $\varphi'_i$.
Accordingly, the partition function given by (\ref{eq:Kphi}) can be
defined on the whole of the $CP^{N-1}$ manifold, if the measure is
invariant under the coordinate change.

Thus, it is clear that the nonlinear model is defined on the whole
manifold by using inhomogeneous coordinates.  Let us remember that, in
the linear model, $\phi_i\ (i=1,\cdots,N-1)$ are related to the
coordinates $\varphi_i$ and then $\phi_N$ is given by the gauge
transformation parameter $\lambda$.  According to (\ref{eq:cdch}), 
to move from $U_0$ to $U_k$, we have only to transform the superfields as
\begin{align}
 \phi'_k=\frac{\phi_N^2}{\phi_k},
~~~\phi'_i=\frac{\phi_N \phi_i}{\phi_k}~~~(i\neq k).
\label{eq:coodtrans}
\end{align}
It is easily seen that the action is unchanged under this transformation.
Consequently, the linear model is also defined on the whole $CP^{N-1}$
manifold.

Here, it should be emphasized that $\phi_N$ is merely a redundant field,
or in other words a compensating field \cite{deWit:1985bn},
which is irrelevant to a coordinates system for $CP^{N-1}$:
$\phi_N=e^{-i\lambda}/g$.  If there is no anomalous term in
(\ref{eq:partfuncf2}), $\phi_i~(i=1,\cdots, N)$ may be interpreted as
homogeneous coordinates for $CP^{N-1}$, and $\phi_N=0$ may represent
hyperplane at infinity in the $CP^{N-1}$.  At the present case, $\phi_N$
is not equal to zero due to a logarithmic singularity of the anomalous
term (\ref{eq:alphaV}). However, this is not a problem for including the
hyperplane at infinity in the model, because the transformation
(\ref{eq:coodtrans}) makes us possible to change coordinate patches and
to include the whole manifold. It is noted that, on the contrary,
the coordinate transformation (\ref{eq:coodtrans}) is breakdown for
$\phi_N=0$.

\subsection{Exact anomalous terms in cut-off theories\label{sec:anomaly}}

The $CP^{N-1}$ model in four dimensions is nonrenormalizable and it
is regarded as a low energy effective field theory with a ultraviolet
cutoff. So, we have to evaluate the anomalous contribution in the gauged
linear model by keeping the cutoff finite. In this section, we consider
the cutoff dependence of the anomalous term by calculating the triangle
diagram.

First, we consider the vacuum functional
\begin{align}
 e^{i\Gamma[V]}=\int d\phi d\phi^\dagger \exp\Big(i\int d^8 z
  K(\phi,\bar{\phi},V)\Big).
\end{align}
Since $U(1)_{\rm local}$ is anomalous, $\Gamma[V]$ is not gauge
invariant due to the triangle diagram. On the other hand, since the
partition 
function (\ref{eq:partfuncf2}) is gauge invariant, the anomaly from the gauge
transformation of $\Gamma[V]$ is canceled by the gauge transformation of
$\alpha[V,\phi_N,\bar{\phi}_N]$:
\begin{align}
\delta \alpha[V,\phi_N,\bar{\phi}_N]=-\delta \Gamma[V].
\label{eq:alphaGamma}
\end{align}
Therefore, $\alpha[V,\phi_N,\bar{\phi}_N]$ can be determined by solving
this equation for given $\delta \Gamma[V]$.

Here let us explain in detail the calculation of $\delta\Gamma[V]$ in
the cutoff theory.
The Lagrangian for the chiral spinor is
given by
\begin{align}
 \int d^2\theta d^2\bar{\theta} \,\phi^\dagger e^{2V} \phi
&=
i\bar{\Psi}\slashed{\partial}P_R\Psi+ v_\mu
\bar{\Psi}\gamma^\mu P_R\Psi+\cdots,
\end{align}
where we have used four-component notation for the spinor, and $v_\mu$
denotes a vector field in $V$. $P_R$ is 
a projection operator on the right-handed fermion field:
$P_R=(1+\gamma_5)/2$. 
The famous two triangle diagrams contribute to the three-point vertex
function of $v^\mu$ \cite{rf:Weinberg}:
\begin{align}
\Gamma^{(3)}_{\mu\nu\rho}(k_1,k_2)&\equiv -N
\int\frac{d^4 k}{i(2\pi)^4} \left\{{\rm
 tr}\left[
\frac{1+\gamma_5}{2}
\frac{1}{-\slashed{k}-\slashed{a}}
\gamma_\mu
\frac{1}{-\slashed{k}-\slashed{a}+\slashed{k}_1}
\gamma_\nu
\frac{1}{-\slashed{k}-\slashed{a}-\slashed{k_2}}
\gamma_\rho
\right]\right.
\nn
&
~~~~~~~+\left.{\rm
 tr}\left[
\frac{1+\gamma_5}{2}
\frac{1}{-\slashed{k}+\slashed{a}}
\gamma_\rho
\frac{1}{-\slashed{k}+\slashed{a}+\slashed{k}_2}
\gamma_\nu
\frac{1}{-\slashed{k}+\slashed{a}-\slashed{k}_1}
\gamma_\mu
\right]\right\},
\label{eq:3vertex}
\end{align}
where $N$ component fermions yield the factor $N$.
As in the NJL model, this integral is divergent
and so we introduce the ultra-violet cutoff $\Lambda$ after Wick
rotation. It is noted that the cutoff is different from the previous one
in (\ref{eq:alphaV}) and there is no simple relation between them.
The four-vector $a^\mu$ is introduced due to arbitrariness of the
momenta carried by internal lines.

More precisely, we can introduce two
four-vectors $a^\mu$ and $b^\mu$ independently to each triangle diagram.
In this case, we have to choose $a^\mu=-b^\mu$ for avoiding non-chiral
anomalies for all three currents as explained in \cite{rf:Weinberg}.
Actually, the charge conjugation matrix $C$ satisfies
$C^{-1}\gamma^\mu C=-{\gamma^\mu}^T$ and we have
\begin{align}
{\rm
 tr}\left[\frac{1}{-\slashed{k}-\slashed{a}}
\gamma_\mu
\frac{1}{-\slashed{k}-\slashed{a}+\slashed{k}_1}
\gamma_\nu
\frac{1}{-\slashed{k}-\slashed{a}-\slashed{k_2}}
\gamma_\rho
\right]
=-{\rm
 tr}\left[
\frac{1}{\slashed{k}+\slashed{a}}
\gamma_\rho
\frac{1}{\slashed{k}+\slashed{a}+\slashed{k}_2}
\gamma_\nu
\frac{1}{\slashed{k}+\slashed{a}-\slashed{k}_1}
\gamma_\mu
\right].
\nonumber
\end{align}
So, the traces which contain no $\gamma_5$ in (\ref{eq:3vertex})
cancel to each other if a momentum
variable is flipped in one diagram: $k^\mu\rightarrow -k^\mu$.
Therefore, only the traces involving $\gamma_5$ are left and
this justifies a choice of $a^\mu=-b^\mu$.

Now, we evaluate the anomaly term $\delta \Gamma[V]$, which corresponds
to the Fourier transformation of the divergence of
(\ref{eq:3vertex}):\footnote{In general, a simple momentum cut-off
breaks gauge invariance and this is a well-known problem, for example,
as seen in dealing with vector mesons in the NJL model \cite{rf:NJL}. In
the NJL model, a conventional gauge invariant form of vertex functions
was used to avoid an ambiguity of mass subtraction.  There are many
other prescriptions proposed to deal with gauge invariance in cut-off
theories. Here, we use arbitrariness in the choice of the momentum shift
in the loop integral in order to ensure gauge invariance. As an
alternative, you may define the model in the gauge invariant way by
higher derivative kinetic term as in \cite{Hamazaki:1994rf}.  In any
case, qualitative features are unchanged.}
\begin{align}
&(k_1+k_2)^\nu \Gamma^{(3)}_{\mu\nu\rho}(k_1,k_2)
\nn
&=
4Ni\epsilon_{\nu\mu\lambda\rho}
\int_{k^2\leq \Lambda^2}
\frac{d^4 k}{(2\pi)^4} 
\left\{
\frac{(k+a)^\nu{k_2}^\lambda}{(k+a)^2(k+a+k_2)^2}
-
\frac{-(k+a)^\nu{k_1}^\lambda}{(k+a)^2(k+a-k_1)^2}
\right\}.
\end{align}
These integrals can be calculated straightforwardly by picking up
anti-symmetric parts on the two indices $\nu$, $\lambda$.
Combining the denominator by the Feynman parameter technique, we perform
the $k$ integration by using the formula in the appendix.
Then if one rotates back to the Minkowski space, the resulting
function is given by
\begin{align}
&
i(k_1+k_2)^\nu \Gamma^{(3)}_{\mu\nu\rho}(k_1,k_2)
\nn
&=
-\frac{N}{8\pi^2} \epsilon_{\nu\mu\lambda\rho}\int_0^1 dx
\left\{
a^\nu {k_2}^\lambda
\,g(-(a+xk_2)^2,-a^2-2xa\cdot k_2-x k_2^2)
\right.
\nn
&
~~~~~~~~~\left.
+a^\nu {k_1}^\lambda
\,g(-(a-xk_1)^2,-a^2+2xa\cdot k_1-x k_1^2)
\right\},
\end{align}
where $g(p^2,m^2)$ is defined by (\ref{eq:gpm}).
This is the exact result for the anomalous vertex function in the
cut-off theory.

Suppose that the currents for the $\mu$, $\rho$ directions are
conserved, we have to choose
$a=k_1-k_2$ as explained in \cite{rf:Weinberg}:
\begin{align}
i(k_1+k_2)^\nu \Gamma^{(3)}_{\mu\nu\rho}(k_1,k_2)
&=
-\frac{N}{4\pi^2} \epsilon_{\nu\mu\lambda\rho}{k_1}^\nu
{k_2}^\lambda\,f(k_1,\,k_2),
\end{align}
where $f(k_1,\,k_2)$ is given by
\begin{align}
f(k_1,\,k_2)&=\frac{1}{2} 
\int_0^1 dx
\left\{
\,g(\,-(k_1-(1-x)k_2)^2,\,-k_1^2+2(1-x)k_1\cdot k_2-(1-x) k_2^2)
\right.
\nn
&
~~~~~~~~~\left.
\,+g(\,-((1-x)k_1-k_2)^2,\,-(1-x){k_1}^2+2(1-x)k_1\cdot k_2-k_2^2)
\right\}.
\label{eq:fdef}
\end{align}
This result is expressed in terms of the chiral current
$J^\mu\equiv \bar{\psi}\bar{\sigma}^\mu \psi
=\bar{\Psi}\gamma^\mu P_R \Psi$:
\begin{align}
 \partial_\nu \left<J^\nu(x)\right>
&=-\frac{N}{32\pi^2}\epsilon_{\nu\mu\lambda\rho}
F^{\nu\mu}
\,f\left(-i\frac{\overleftarrow{\partial}}{\partial x},\,
-i\frac{\overrightarrow{\partial}}{\partial
  x}\right) F^{\lambda\rho}.
\end{align}
The expansion in powers of $1/\Lambda$ is evaluated as
\begin{align}
  \partial_\nu \left<J^\nu(x)\right>
&=-\frac{N}{32\pi^2}\epsilon_{\nu\mu\lambda\rho}
F^{\nu\mu}F^{\lambda\rho}
+\frac{N}{96\pi^2\Lambda^2}\epsilon_{\nu\mu\lambda\rho}
F^{\nu\mu}\Box F^{\lambda\rho}+O(1/\Lambda^4),
\end{align}
where the first term agrees with the conventional chiral
anomaly.\footnote{According to calculation in the section 22 of
\cite{rf:Weinberg}, correction terms of order $1/\Lambda^2$ are
naturally appeared in the anomalous term as far as we keep the cutoff
finite. ($\Lambda$ should be regarded as a radius $P$ of a large three
sphere in the book \cite{rf:Weinberg}.)  Also by applying the Fujikawa
method \cite{KS}, it is easily seen that the correction term appears in
a Jacobian factor for a finite cutoff.  Then, it is interesting to
understand how to deal with the index theorem in cutoff theories, but
this is out of scope in this paper.}

Since the operator $f$ consists of space-time derivatives, we can
easily provide $\delta \Gamma[V]$ in the supersymmetric
model. Finally, from $\delta \Gamma[V]$ and (\ref{eq:alphaGamma}),
the resulting anomalous term can be obtained as
\begin{align}
 \alpha[V,\phi_N,\bar{\phi}_N]
=-\frac{N}{16\pi^2}\int d^4 x d^2\theta 
\log (g\,\phi_N)\,W^\alpha
f\left(-i\frac{\overleftarrow{\partial}}{\partial x},\,
-i\frac{\overrightarrow{\partial}}{\partial
  x}\right) W_\alpha+{\rm h.c.}.
\label{eq:exactAlphaV1}
\end{align}
This is an exact result for (\ref{eq:alphaV}) including all orders of
$\Lambda$.

\subsection{Effective potentials in the $1/N$ leading order
\label{sec:CPNpot}}

Now that the $CP^{N-1}$ model is formulated as the consistent linear
model, we can consider the effective potential of this model in the
$1/N$ expansion.
In the Wess-Zumino gauge, 
the scalar components are
the D-term $-D$ of the vector superfield $V$ 
and the first component of $\phi_N$.
As in \cite{Kugo:2017qma}, we take
negative sign convention for the D-term of $V$.
The F-term of $\phi_N$ are irrelevant to the effective potential.

In order to perform the $1/N$ expansion, we define the coupling $g^2$ by
\begin{align}
 g^2\equiv \frac{G}{N},
\end{align}
and we study the limit of large $N$ with fixed $G$.
This is a conventional choice used in the $CP^{N-1}$ model.
Moreover, since
$g \phi_N$ should be order one for the anomalous term to be
leading order, the vacuum expectation value of $\phi_N$
should be defined as
\begin{align}
 \langle \phi_N\rangle\equiv \sqrt{N} z,
\end{align}
where $z$ is a fixed complex number in the $1/N$ expansion.

Substituting these component fields to (\ref{eq:exactAlphaV1}), we can
calculate an anomalous contribution to the effective action:
\begin{align}
 \alpha[V,\phi_N,\bar{\phi}_N]=-\frac{N}{16\pi^2}\int d^4 x \log(G|z|^2)\,
D\,f\left(-i\frac{\overleftarrow{\partial}}{\partial x},\,
-i\frac{\overrightarrow{\partial}}{\partial
  x}\right) D.
\label{eq:AnomalyGamma}
\end{align}
For constant $D$, the operator $f$ becomes one and so a quadratic term
of $D$ is generated in the effective potential.

We notice that  
for constant $W_\alpha$,
higher order correction terms
may arise from other diagrams (square, pentagon and so on)
in the the superpotential as
\begin{align}
 \log(g\phi_N) \,\Lambda^3 F\left(\frac{W^\alpha
  W_\alpha}{\Lambda^3}
\right),
\end{align}
where $F(\cdots)$ denotes a certain function.
If we expand it in the power series of $W^\alpha
W_\alpha/\Lambda^3$,
since the constant fields
are included as $W^\alpha W_\alpha=\theta \theta D^2+\dots$ and
$\log(g\phi_N)=\log(G z)+\cdots$, the quadratic and higher powers does not
contribute to the effective potential. So, (\ref{eq:AnomalyGamma}) leads
to an exact result of the anomalous effective potential.

Consequently, we can provide the effective potential in the leading
order in the $1/N$ expansion:
\begin{align}
 \frac{1}{N}V(z,D)&=-\frac{1}{G} D+D|z|^2
+\frac{1}{16\pi^2}D^2 \log(G|z|^2)
\nn
&+\frac{1}{32\pi^2}
\left[\Lambda^4 \log\left(1+\frac{D}{\Lambda^2}\right)
- D^2\log \left(1+\frac{\Lambda^2}{D}\right)+D\Lambda^2\right].
\label{eq:potCPN}
\end{align}
Here, the first and second terms arise from the tree level action, where
we note again the negative sign convention of the D-term. The third term is
the anomalous potential from (\ref{eq:AnomalyGamma}). The forth term is
given by one-loop calculation, which is performed 
in a supersymmetric NJL model in \cite{Kugo:2017qma}. In the calculation,
$D$ is a mass square parameter for the scalar component of $\phi$ and so
$D$ must be positive for a consistent vacuum.

The stationarity condition with respect to $z$ is
\begin{align}
 \frac{\delta V}{\delta z}=
\frac{D}{z}\,\Big( \frac{1}{16\pi^2} 
D+|z|^2\Big)=0.
\label{eq:dVdz}
\end{align}
Then, we conclude $D=0$ and so supersymmetry is never broken in the
leading order.

Another stationarity condition leads to
\begin{align}
 \frac{\delta V}{\delta D}=0
&~\Rightarrow~
-\frac{1}{G}+|z|^2+ \frac{1}{32\pi^2}\left[
 2\Lambda^2-2D\log \left(1+\frac{\Lambda^2}{D}\right)
+4D\log(G|z|^2)\right]
=0.
\label{eq:dVdD}
\end{align}
Substituting $D=0$ into the above, we find
\begin{align}
 |z|^2=\frac{1}{G}-\frac{\Lambda^2}{16\pi^2}
\label{eq:zvev}
\end{align}
The model becomes inconsistent if $G$ is larger than
$G_{\rm cr}=16\pi^2/\Lambda^2$.

Accordingly, we conclude that, in the $1/N$ leading order, the model has
a stable vacuum only for the weak coupling $G<G_{\rm cr}$, and
supersymmetry is unbroken in this vacuum.

Here it should be noted that the anomalous potential has an important
role on the robustness of supersymmetry. If we naively quantize
the gauged linear model without the anomalous term, the
stationarity condition with respect to $z$ becomes $D z^*=0$ instead of
(\ref{eq:dVdz}) and so we have $D=0$ or $z=0$.
The stationarity condition with respect to $D$
implies the gap equation
\begin{align}
 |z|^2-\int 
\frac{d^4 k}{(2\pi)^4}\left(\frac{1}{k^2}-\frac{1}{k^2+D}\right)=\frac{1}{G}
-\frac{1}{G_{\rm  cr}},
\end{align}
which is the same as that of a bosonic $CP^{N-1}$
model\cite{Bando:1987br}. Then, we might have two phases: (i) $G<G_{\rm
cr}$, $|z|\neq 0,\ D=0$ and (ii) $G>G_{\rm cr}$, $|z|=0,\ D\neq 0$. While
the first case corresponds to the above supersymmetric model, the second
is appeared as a new phase.
If there were no anomaly, $D$ would acquire a vacuum expectation value in
strong coupling region and so
supersymmetry would be spontaneously broken.
But it is not the case and so it is regarded that the anomalous term
keeps supersymmetry unbroken. 

We note that, although there is no vacuum in the strong coupling
region in the $1/N$ leading order,
there still remains a possibility of finding a vacuum in higher order or
by considering some nonperturbative effects.

\subsection{Dynamical vector supermultiplets
\label{sec:vector}}

We showed that $z$ has the vacuum expectation
value (\ref{eq:zvev}) in the weak coupling region.
On this vacuum, the anomalous term
(\ref{eq:exactAlphaV}) induces the kinetic term for the vector
superfield:
\begin{align}
&-\frac{N}{16\pi^2}\int d^4 x d^2\theta 
\,\log(\sqrt{G} z)\,W^\alpha
f\left(-i\frac{\overleftarrow{\partial}}{\partial x},\,
-i\frac{\overrightarrow{\partial}}{\partial
  x}\right) W_\alpha+{\rm h.c.}.
\label{eq:exactAlphaV}
\end{align}
It is well-known that, in general, vector bosons are
dynamically generated in the model with hidden local symmetries
\cite{Bando:1987br}. Also in this model,  loop
diagrams of components of $\phi$ generate the kinetic term for a vector
boson. In addition, the
anomalous term (\ref{eq:exactAlphaV}) supplies the kinetic term, 
which however enhances the possibility of wrong sign due to the
logarithmic function. 
If the logarithmic function is positive, the anomalous term encourages
the appearance of negative metric states.

Fortunately, it can be easily seen that
the large $N$ dynamics prohibits such a negative metric state.
For the vacuum expectation value (\ref{eq:zvev}), we find
\begin{align}
 G|z|^2=1-\frac{G}{G_{\rm cr}}<1~~~(\ G<G_{\rm cr}\ ).
\end{align}
Therefore, the kinetic term of the vector superfield is well behaved
since the logarithmic function becomes negative for $G<G_{\rm cr}$.
Then,
the anomalous term (\ref{eq:exactAlphaV}) leads to the vertex
function of the vector field:
\begin{align}
 \Gamma^{{\rm A}(2)}_{\mu\nu}(p)=(p^2\eta_{\mu\nu}-p_\mu p_\nu)
\frac{N}{32\pi^2}\log\Big(1-\frac{G}{G_{\rm cr}}\Big)\,f(p,-p),
\label{eq:vecGamma2}
\end{align}
where $f(p,-p)$ can be evaluated explicitly from (\ref{eq:fdef}).
For $p^2>0$, we find
\begin{align}
 f(p,-p)=\frac{1+7 p^2/3\Lambda^2}{1+2p^2/\Lambda^2}.
\end{align}

Now, 
we calculate all of the two-point vertex
function of the vector field for the time-like momentum.
For loop integrations
with a cut-off, we
have the freedom to choose a momentum shift carried by internal lines,
as well as the anomaly calculation in the section
\ref{sec:anomaly}. Here, by adopting a symmetric momentum shift
($a^\mu=-p^\mu/2$), the
vertex function $\Gamma^{'(2)}_{\mu\nu}(p)$
for the vector component is given by
\begin{align}
&
 \Gamma^{'(2)}_{\mu\nu}(p)=\Gamma^{{\rm f}(2)}_{\mu\nu}(p)
+\Gamma^{{\rm b}(2)}_{\mu\nu}(p),
\\
&
 \Gamma^{{\rm f}(2)}_{\mu\nu}(p)=
-N \int_0^1 dx\,
\int_{k^2\leq \Lambda^2} \frac{d^4 k}{(2\pi)^4}
\frac{4 k_\mu k_\nu
-p_\mu p_\nu
-2(k^2-p^2/4)\eta_{\mu\nu}}{
\{k^2+2(1/2-x)p\cdot k+p^2/4\}^2},
\\
&
 \Gamma^{{\rm b}(2)}_{\mu\nu}(p)=
N \int_0^1 dx\,
\int_{k^2\leq \Lambda^2} \frac{d^4 k}{(2\pi)^4}
\frac{4 k_\mu k_\nu
-2(k^2+p^2/4)\eta_{\mu\nu}}{
\{k^2+2(1/2-x)p\cdot k+p^2/4\}^2},
\end{align}
where $\Gamma^{{\rm f}}$ and $\Gamma^{\rm b}$ are coming from fermion
and boson one-loop diagrams, respectively. 
After the $k$ integration by using the formula in the appendix, we
find that each vertex function
includes a quadratic term of $\Lambda$, which
corresponds to the vector self-energy.
It implies that gauge symmetry is broken by introducing the cut-off
parameter. However, the quadratic terms cancel to each other 
in the total vertex function owing to supersymmetry.
As a result, the vertex function is expressed in the conventional gauge
invariant 
form: for $0<p^2<4\Lambda^2$ in the Minkowski space,
\begin{align}
 \Gamma_{\mu\nu}^{'(2)}(p)
&=
-(p^2\eta_{\mu\nu}-p_\mu p_\nu)
\frac{N}{16\pi^2}\Big(1
+\log\frac{4\Lambda^2-p^2}{4p^2}+i\pi\Big).
\end{align}
The integral is calculated as a real number in the Euclidean space, but the
imaginary part appears in the Minkowski
space due to the logarithm function. 

Combining these results with tree level terms, the resulting vertex
function for the time-like momentum is given by
\begin{align}
 \Gamma^{(2)}_{\mu\nu}(p)&=
-(p^2\eta_{\mu\nu}-p_\mu p_\nu)F(p^2)
+m^2\eta_{\mu\nu},
\\
F(p^2)&=\frac{N}{16\pi^2}
\Big\{1+\log\frac{4-p^2/\Lambda^2}{4p^2/\Lambda^2}
-\frac{1+7 p^2/3\Lambda^2}{2(1+2p^2/\Lambda^2)}
\log\Big(1-\frac{G}{G_{\rm cr}}\Big)\Big\}+i \frac{N}{16\pi},
\label{eq:Fp2}
\\
m^2&=
N\Big(\frac{2}{G}-\frac{2}{G_{\rm cr}}\Big).
\label{eq:m2}
\end{align}

From this vertex function, we could expect  that a massive vector
particle appears dynamically, however it includes the nonzero imaginary
part and so the ``would be'' vector particle is unstable.
Actually, the vector particle has couplings with the scalar and
spinor components of $\phi$, which remain massless in the $1/N$ leading
order and so the vector state decays into these massless particles.

\begin{figure}[t]
\centerline{
\includegraphics[width=11cm]{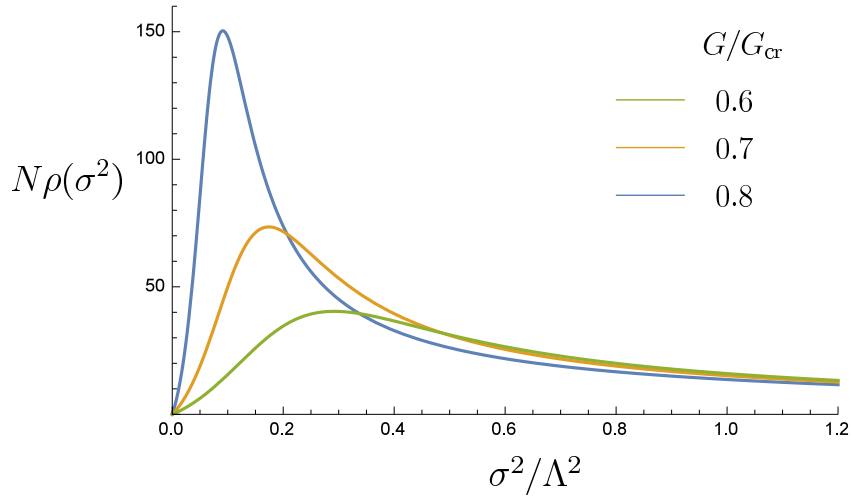}
}
\caption{The plots of the spectral function for the unstable vector state.}
\label{fig:specfunc}
\end{figure}

Finally, we elucidate the behavior of the unstable vector state in terms
of the spectral function. 
The propagator can be derived from the vertex function
(\ref{eq:vecGamma2}):
\begin{align}
 \Delta_{\mu\nu}(p)&=i\Delta'(p)\,\left\{
\eta_{\mu\nu}-\frac{p_\mu
p_\nu}{m^2}
F(p^2)\right\},~~~
\Delta'(p)=\frac{1/F(p^2)}{m^2/F(p^2)-p^2}.
\end{align}
Here we forget for a moment that $F(p^2)$ is divergent for
$\Lambda^2\rightarrow \infty$ as in \cite{rf:NJL}. If so, the spectral
function $\rho(\sigma^2)$ is given by the imaginary part of $\Delta'(p)$
and then $\Delta'(p)$ is expressed by $\rho(\sigma^2)$:
\begin{align}
 \Delta'(p)=\int_0^{\Lambda^2} d\sigma^2\,
\frac{\rho(\sigma^2)}{\sigma^2-p^2-i\epsilon},
\end{align}
where a new cutoff is introduced as in \cite{rf:NJL}, although there are
no simple relation between both cutoffs.
By using (\ref{eq:Fp2}) and (\ref{eq:m2}), we can evaluate
$\rho(\sigma^2)$ numerically and the resulting plots are depicted in
Fig.~\ref{fig:specfunc}.
We note that $\rho(\sigma^2)$ is given by order $1/N$.

From these plots, we find a peak in the region
$\sigma^2\lesssim \Lambda^2$ for the coupling
$G\gtrsim 0.5\, G_{\rm cr}$,
but the width is 
large and the peak is hard to distinguish for $0.7\,G_{\rm cr}\gtrsim
G\gtrsim 0.5\,G_{\rm cr}$.
Near the critical coupling, the position of the peak
approaches to $\sigma^2\sim 0$ and the width becomes gradually narrower.

\begin{figure}[t]
\centerline{
\includegraphics[width=9cm]{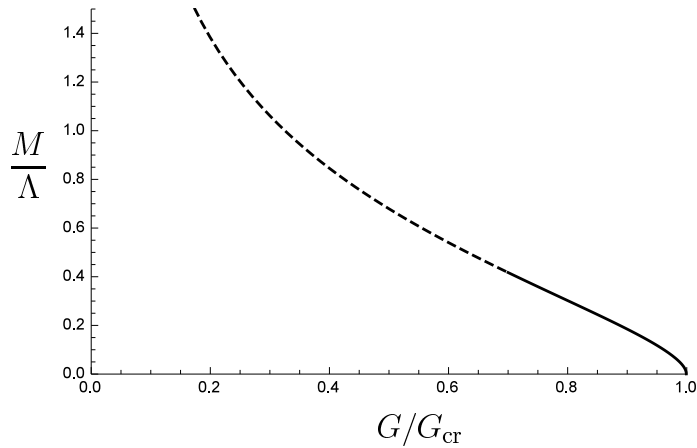}
}
\caption{The plots of the position of the peak of $\rho(\sigma^2)$,
 $\sigma=M$.
It corresponds to the mass of the unstable vector multiplet.}
\label{fig:vecmass}
\end{figure}

The position of the peak can be evaluated numerically
by using the numerical results of $\rho(\sigma^2)$.
The resulting plots are shown
in Fig.~\ref{fig:vecmass}.
We find that the ``mass'' of the unstable vector state decreases to zero
for the coupling $G$ approaching to $G_{\rm cr}$.
Since supersymmetry is not broken in this vacuum,
the vector supermultiplet is dynamically
generated for $G\gtrsim 0.7\,G_{\rm cr}$, but it is
unstable.

Most interestingly, we find that the spectral function approaches
rapidly to a delta function for $G\rightarrow G_{\rm cr}$, namely
$\rho(\sigma^2) \rightarrow Z\,\delta(\sigma^2)$.
This behavior suggests that a massless vector supermultiplet is
dynamically generated and the $U(1)$ gauge symmetry is restored at the
critical coupling. Unfortunately, the analysis just at $G=G_{\rm cr}$
seems to be subtle in the leading order, because $|z|^2$ becomes zero
and so the logarithmic term in the effective potential diverges.

\section{Nonlinear sigma models with F-term constraint}

\subsection{$SO(N)/SO(N-2)\times U(1)$ model
\label{sec:SOmodel}}

We consider a supersymmetric nonlinear sigma model based on the manifold
$SO(N)/SO(N-2)\times U(1)$ \cite{Higashijima:1999ki}. The model is
formulated by a gauged linear
sigma model as well as the $CP^{N-1}$ model. We introduce the chiral
superfields $\phi_i~~(i=1,\cdots, N)$ and the K\"ahler potential is the
same as (\ref{eq:KphiV}). In addition, the linear model has the
superpotential by using an extra chiral superfield:
\begin{align}
 W(\phi_0,\,\phi)=\frac{1}{2}\phi_0\phi^2,
\label{eq:superpot}
\end{align}
where the chiral superfield $\phi_0$ corresponds to a Lagrange
multiplier and then it induces the constraint $\phi^2=0$. For the $U(1)$
symmetry, $\phi$ and $\phi_0$ has the charge $+1$ and $-2$,
respectively.

In order to transform back to the nonlinear model, we have to fix the
 gauge of the $U(1)$ symmetry as $\phi_N=1/g$ similar to the case of the
 $CP^{N-1}$ model.  Here, we should notice that this rewriting also
 suffers from the anomaly. Since the total $U(1)$ charge for $\phi_0$
 and $\phi_i$ equals to $N-2$, the anomalous term turns out to be given
 by\footnote{It is noted that, as in the $CP^{N-1}$ model,
 $\phi_i~(i=1,\cdots, N-1)$ are related to local coordinates of the
 manifold. $\phi_0$ and $\phi_N$ are irrelavant to local
 coordinates. So, we can use coordinate transformations to cover the
 whole of the manifold.}
\begin{align}
 \alpha[V,\phi_N,\bar{\phi}_N]
=-\frac{N-2}{16\pi^2}\int d^4 x d^2\theta 
\log (g\,\phi_N)\,W^\alpha
f\left(-i\frac{\overleftarrow{\partial}}{\partial x},\,
-i\frac{\overrightarrow{\partial}}{\partial
  x}\right) W_\alpha+{\rm h.c.}.
\label{eq:exactAlphaV2}
\end{align}
As a result, 
the symmetry of the action is reduced to $SO(N-1)\times U(1)_{\rm
local}$, while the K\"ahler potential has the symmetry $SO(N)\times
U(1)_{\rm local}$.

In the background $\langle \phi_0\rangle
=[w,\,0,\,h]$,
the part of the Lagrangian derived from (\ref{eq:superpot}) is expanded
by the component fields $\phi^i=[A^i,\,\psi^i,\,F^i]$ as
\begin{align}
 \int d^2\theta\,W(\phi_0,\,\phi)+{\rm h.c.}=
w F^i A^i+\frac{1}{2}h A^iA^i-w\psi^i\psi^i+{\rm h.c.}.
\label{eq:Wterm}
\end{align}
Eliminating the auxiliary fields $F_i$ by the equations of motion
${F^i}^*+w A^i=0$, (\ref{eq:Wterm}) yields mass terms for component
fields. By including the contribution from the K\"ahler potential,
the mass terms in this background are given as
\begin{align}
 {\cal L}_{\rm mass}=-(D+|w|^2){A^i}^\dagger A^i +\left(
\frac{1}{2}hA^iA^i-w\psi^i\psi^i+{\rm h.c.}\right).
\label{eq:massterm}
\end{align}

\subsection{Effective potentials including F-terms
\label{sec:SOpot}}

The mass term (\ref{eq:massterm}) is essentially same as that of the
supersymmetric NJL model analyzed in
\cite{Kugo:2017qma}. For the 
scalar, the mass square eigenvalues are given by $D+|w|^2\pm |h|$.
According to \cite{Kugo:2017qma}, 
the effective
potential in the $1/N$ leading order can be calculated as
\begin{align}
\frac{1}{N}V(z,\,D,\,w,\,h)&=-\frac{1}{g^2}D+N(D+|w|^2-|h|\cos\theta)|z|^2
+\frac{1}{16\pi^2}D^2 \log(G|z|^2)
\nn
&+\frac{1}{16\pi^2}
 \Big\{F(D+|w|^2+|h|)+F(D+|w|^2-|h|)-2 F(|w|^2)
\Big\},
\label{eq:potSO}
\end{align}
where $\theta$ is the phase of $hA^i A^i$ and
the function $F(x)$ is defined by
\begin{align}
 F(x)=\frac{1}{2}
\Big[\log(1+x)-x^2 \log\Big(1+\frac{1}{x}\Big)+x\Big].
\end{align}
We set the cutoff $\Lambda$ equal to one for simplicity.
The potential (\ref{eq:potSO}) reduces to a similar expression
to the previous one
(\ref{eq:potCPN}) if taking the limit $h,\,w\rightarrow 0$.
We note that the factor of the anomalous term $N-2$ is approximated as
$N$ for large $N$.

Differentiating the potential (\ref{eq:potSO}), the stationarity
conditions are given by 
\begin{align}
 \frac{\delta V}{\delta \theta}=0
&~\Rightarrow~
|h||z|^2\sin\theta=0,
\label{eq:dV1}
\\
 \frac{\delta V}{\delta |h|}=0
&~\Rightarrow~
I(D+|w|^2+|h|)-I(D+|w|^2-|h|)
=16\pi^2 |z|^2\cos\theta,
\label{eq:dV2}
\\
 \frac{\delta V}{\delta w}=0
&~\Rightarrow~
w^* \Big\{I(D+|w|^2+|h|)+I(D+|w|^2-|h|)
\Big\}=-16\pi^2 w^* |z|^2,
\label{eq:dV3}
\\
 \frac{\delta V}{\delta z}=0
&~\Rightarrow~
\frac{1}{z}\,\Big( \frac{1}{16\pi^2} 
D^2+(D-|h|\cos\theta )|z|^2\Big)=0,
\label{eq:dV4}
\\
 \frac{\delta V}{\delta D}=0
&~\Rightarrow~
 -\frac{1}{G}+|z|^2+\frac{1}{8\pi^2} D \log(G|z|^2)
\nn
&~~~~~~~
+\frac{1}{16\pi^2} \Big\{I(D+|w|^2+|h|)+I(D+|w|^2-|h|)
\Big\}=0,
\label{eq:dV5}
\end{align}
where $I(x)$ is defined by
\begin{align}
 I(x)\equiv F'(x)=1-x\log\Big(1+\frac{1}{x}\Big).
\end{align}

The stationarity condition (\ref{eq:dV1}) implies that $\theta=0\ {\rm or
}\ \pi$, or $|h|=0$. Note that $|z|$ must not be zero since the
potential includes $\log |z|$.
Since $I(x)$ is a monotonically decreasing function\cite{Kugo:2017qma},
we find, if $|h|\neq 0$, 
\begin{align}
 I(D+|w|^2+|h|)-I(D+|w|^2-|h|)<0.
\end{align}
So, from (\ref{eq:dV2}), it follows that $\theta=\pi$ if
$|h|\neq 0$. However, these values do not satisfy the stationarity
condition (\ref{eq:dV5}) and so $|h|$ must be zero.
Then, from
(\ref{eq:dV2}) and (\ref{eq:dV4}), it follows that
$\theta$ must be $\pi/2$
and $D$ must be zero.
At this stage, we conclude that supersymmetry is unbroken in this model
since $D=0$ and $h=0$.

From (\ref{eq:dV3}) and (\ref{eq:dV5}), we find that if $w\neq 0$, 
\begin{align}
  -\frac{1}{G}+\frac{1}{8\pi^2} D \log(G|z|^2)=0.
\end{align}
It is inconsistent for $D=0$ and so $w$ must be zero.

After all, $D$, $h$ and $w$ are zero, and $|z|$ is given by the same
expression of (\ref{eq:zvev}). At this vacuum, the effective action is
essentially same as that of the $CP^{N-1}$ model in the $1/N$ leading
order.
Therefore, the analysis of
the vector boson is also the same and so one massive vector particle
appears in this model, but it decays to massless components.

\section{Concluding Remarks
\label{sec:conclu}}

We have shown that the supersymmetric $CP^{N-1}$ and
$SO(N)/SO(N-1)\times U(1)$ models are formulated as anomalous gauge
theories.  By the anomalous term, the gauged linear models have smaller
symmetries of the action than conventional ones: the remaining symmetry
is $SU(N-1)_{\rm global}\times U(1)_{\rm local}$ for $CP^{N-1}$, and
$SO(N-1)_{\rm global}\times U(1)_{\rm local}$ for $SO(N)/SO(N-2)_{\rm
global}\times U(1)_{\rm local}$.

In the $1/N$ leading order, the linear model has a vacuum for $G<G_{\rm
cr}$, where the $U(1)_{\rm local}$ symmetry is broken but supersymmetry
is unbroken. It is a remarkable feature of both models that there is no
stable vacuum for $G>G_{\rm cr}$ in the $1/N$ leading order.

From the analysis of the spectral function, we
expect that the dynamical gauge boson becomes massless at the critical
coupling and so the $U(1)_{\rm local}$ symmetry is restored.
To show this, it is necessary to study the models in the
strong coupling regime by
other methods than the $1/N$ leading order.
In particular, it is interesting to clarify the fate of supersymmetry
for $G>G_{\rm cr}$.

It has been shown that all supersymmetric nonlinear sigma models for the
hermitian symmetric space are formulated as gauge
theories, although the anomaly is not included in
\cite{Higashijima:1999ki}. In this 
paper, we deal with the two models for the
hermitian symmetric space
and show that the anomaly should
be taken into account in the models. Then, it is natural to ask whether
the anomalous term is required for analyzing the model for other
hermitian symmetric space.

In the case of the Grassmann manifold $G_{M,N}$, the linear model is
described by a chiral superfield of the $(N,\,\bar{M})$ representation
of $U(N)_L\times U(M)_R$ and the model has no F-term constraint. Since
$U(M)_R$ is gauged in this model, the anomalous term should be added in
the nonlinear sigma model for $G_{M,N}$.

For $Sp(N)/U(N)$ and $SO(2N)/U(N)$, we have similarly
a chiral superfield $\phi$ and an additional chiral field $\phi_0$ to
impose the F-term constraint. Although the gauge symmetry is
non-abelian, it can be easily 
seen that the anomalous term is required also in this case by
considering $U(1)_{\rm D}$, which is a subgroup of $U(N)$
\cite{Higashijima:1999ki}. For $U(1)_{\rm D}$, $\phi$ and $\phi_0$ have
$1$ and $-2$ charge, respectively. Counting the total charge,
the anomalous factor for $U(1)_{\rm D}$ is given by
$N(N+1)$ for $Sp(N)/U(N)$, and $N(N-1)$
for $SO(2N)/U(N)$. Since these factors are nonzero, we
should include the anomalous term in the linear model for these target
manifolds.

Similarly, we can deal with $E_6/SO(10)\times U(1)$ and $E_7/E_6\times
U(1)$ in terms of the $U(1)_{\rm D}$ charge. In the case of
$E_6/SO(10)\times U(1)$, there are two chiral superfields of the $27$
representation of $E_6$ and they have $1$ and $-2$ charge. So, we need
the anomalous term in the linear model. For $E_7/E_6\times U(1)$, we
have two chiral superfields of the 56 representation of $E_7$, which
have $1$ and $-3$ charge for $U(1)_D$ and so the anomalous term is
required. Consequently, we conclude that it is necessary to include the
anomalous term in all linear models corresponding to the nonlinear
sigma model whose target manifold is the hermitian symmetric space. 

Finally, we comment on a supersymmetric NJL model proposed by Cheng,
Dai, Faisei and Kong\cite{Cheng:2015dgt,Cheng:2016dvq}.
The model is given by the K\"ahler potential truncating higher-order
terms of (\ref{eq:Kphiseries}). One analysis of
the model was performed 
in \cite{Kugo:2017qma} by introducing an auxiliary vector superfield
and calculating an effective potential in the $1/N$ leading order.
Relating to an auxiliary vector superfield, the model has hidden $U(1)$
local symmetry with the anomaly, as well as in the $CP^{N-1}$ model.
However, the anomalous term was not included in the effective potential
in the previous analysis. The result including the anomaly will be
reported in the near future\cite{rf:KOT}.

\section*{Acknowledgments}
The authors would like to thank H.~Itoyama, T.~Kugo, N.~Maru, H.~Ohki
and S.~Seki for valuable discussions.
We also acknowledge to an anonymous referee for their useful comments
and suggestions.
The research of T. T. was
supported in part by Nara Women's University Intramural Grant for
Project Research and JSPS KAKENHI Grant Number JP20K03972.

\appendix
\section{Feynman integrals in cut-off theories
\label{sec:appendix}}

First, let us consider the Feynman integral
\begin{align}
 I=\int_{k^2\leq \Lambda^2}\frac{d^4k}{(2\pi)^4}
\frac{1}{k^2+2k\cdot p+m^2},
\end{align}
where $k^\mu$ and $p^\mu$ are Euclidean momenta.
The dot product for the two momenta is written by
$k\cdot p=|k||p|\cos\theta$,
where $\theta$ is the angle between the two vectors and $|k|$ is the norm.
Writing $k=|k|$ and $p=|p|$, the Feynman integral is expressed as
\begin{align}
I=\frac{4\pi}{16\pi^4}\int_0^\Lambda dk k^3\int_0^\pi d\theta
\frac{\sin^2\theta}{k^2+m^2+2kp \cos\theta} ,
\end{align}
where we have used $d^4 k =dk \,d\theta\,4\pi k^3 \sin^2
\theta$ in four dimensions.

Here, the $\theta$ integration can be performed by the formula,
\begin{align}
 \int_0^\pi \,d\theta \frac{\sin^2\theta}{a+2b
  \cos\theta}=\frac{\pi}{4b^2}
\left(a-\sqrt{(a+2b)(a-2b)}\right)~~~~(\,a>2|b|,~~b\neq 0\,).
\end{align}
In the case of $m>p$, we have $k^2+m^2>2kp$ and so the Feynman integral
becomes
\begin{align}
 I=\frac{1}{16\pi^2}\int_0^\Lambda dk
\frac{k}{p^2}\left\{k^2+m^2-\sqrt{(k^2+m^2+2kp)(k^2+m^2-2kp)}\right\}.
\end{align}
Then, the $k$ integration can be easily performed. The
resulting integral is
\begin{align}
 I&=\frac{1}{16\pi^2}\Big\{\frac{\Lambda^4+\Lambda^2 m^2-\Lambda^2 p^2}{
\Lambda^2+m^2}
+\frac{p^2}{2}\Big(1-\frac{2p^2}{\Lambda^2+m^2}\Big)g(p^2,m^2)
+(p^2-m^2)h(p^2,m^2)\Big\},
\label{eq:int1}
\end{align}
where $h(p^2,m^2)$ and $g(p^2,m^2)$ are defined by
\begin{align}
 g(p^2,m^2)&=\frac{\Lambda^4}{2p^4}
\Big(1+\frac{m^2}{\Lambda^2}\Big)
\left\{1+\frac{m^2}{\Lambda^2}-
\sqrt{\Big(1+\frac{m^2}{\Lambda^2}\Big)^2-\frac{4p^2}{\Lambda^2}}
-\frac{2p^2}{\Lambda^2+m^2}
\right\},
\label{eq:gpm}
\\
h(p^2,m^2)&=\log\frac{\Lambda^2+m^2-2p^2+\sqrt{(\Lambda^2+m^2)^2
-4\Lambda^2 p^2}}{2(m^2-p^2)}.
\end{align}

Next, we illustrate the integration with a momentum in the numerator of
the integrand:
\begin{align}
\int_{k^2\leq \Lambda^2}\frac{d^4k}{(2\pi)^4}
\frac{k_\mu}{(k^2+2k\cdot p+m^2)^2}
&=
-\frac{1}{2}\frac{\partial}{\partial p^\mu}
 \int_{k^2\leq \Lambda^2}\frac{d^4k}{(2\pi)^4}
\frac{1}{k^2+2k\cdot p+m^2}
\nn
&=
\frac{4\pi}{16\pi^4}\int_0^\Lambda dk k^3\int_0^\pi d\theta
\frac{k \sin^2\theta \cos\theta}{(k^2+m^2+2kp \cos\theta)^2}\frac{p_\mu}{p}.
\end{align}
By using the formula
\begin{align}
  \int_0^\pi d\theta \frac{\sin^2\theta \cos\theta}{a+2b \cos \theta}
=
\frac{\pi(-a^2+2b^2)}{8b^3}+\frac{\pi a}{8b^3}\sqrt{(a+2b)(a-2b)},
\end{align}
the $\theta$ integration is performed and then we find that the result
of the $k$ integration is given by
\begin{align}
\int_{k^2\leq \Lambda^2}\frac{d^4k}{(2\pi)^4}
\frac{k_\mu}{(k^2+2k\cdot p+m^2)^2}
=\frac{p_\mu}{16\pi^2}\Big\{\frac{\Lambda^2}{
\Lambda^2+m^2}
+\frac{1}{2}\Big(1+\frac{2p^2}{\Lambda^2+m^2}\Big)
g(p^2,m^2)-h(p^2,m^2)\Big\}. 
\label{eq:int2}
\end{align}

Other Feynman integrals can be calculated by similar procedure.
We give the results of calculation of other Feynman integrals used in this
paper:
\begin{align}
&
\int_{k^2\leq \Lambda^2}\frac{d^4k}{(2\pi)^4}
\frac{1}{(k^2+2k\cdot p+m^2)^2}
=\frac{1}{16\pi^2}\Big\{-\frac{\Lambda^2}{
\Lambda^2+m^2}
-\frac{p^2}{\Lambda^2+m^2}\,g(p^2,m^2)+h(p^2,m^2)\Big\},
\label{eq:int3}
\end{align}
\begin{align}
&
\int_{k^2\leq \Lambda^2}\frac{d^4k}{(2\pi)^4}
\frac{k_\mu k_\nu}{(k^2+2k\cdot p+m^2)^2}
\nn
&
=\frac{1}{16\pi^2}\frac{p_\mu p_\nu}{p^2}\Big\{
\frac{\Lambda^2(\Lambda^2+m^2-3p^2)}{
2(\Lambda^2+m^2)}
-\frac{1}{4}\Big(\Lambda^2+m^2+p^2
+\frac{6p^4}{\Lambda^2+m^2}\Big)\,g(p^2,m^2)
\nn
&~~~~~~~~~~~~~~~~~~~~+\frac{3p^2-m^2}{2}h(p^2,m^2)\Big\}
\nn
&
~~+\frac{1}{16\pi^2}\frac{-1}{2}\left(\delta_{\mu\nu}
-\frac{p_\mu p_\nu}{p^2}\right)\Big\{
-\frac{\Lambda^2(\Lambda^2+3 m^2-3p^2)}{
3(\Lambda^2+m^2)}
\nn
&
~~~~~~~~~~~~~~~~~~~~~~~~~~
-\frac{1}{6}\Big(\Lambda^2+m^2-p^2
+\frac{(4m^2-6p^2)p^2}{\Lambda^2+m^2}\Big)\,g(p^2,m^2)
-(p^2-m^2)h(p^2,m^2)\Big\}.
\label{eq:int4}
\end{align}

It is noted that the consistency of (\ref{eq:int1}), (\ref{eq:int2}),
(\ref{eq:int3}) and (\ref{eq:int4}) can be checked by the relation
\begin{align}
 \frac{1}{k^2+2k\cdot p+m^2}
= \delta^{\mu\nu}\frac{k_\mu k_\nu}{(k^2+2k\cdot p+m^2)^2}
+
2p^\mu \frac{k_\mu}{(k^2+2k\cdot p+m^2)^2}
+
m^2\frac{1}{(k^2+2k\cdot p+m^2)^2}.
\end{align} 


\end{document}